\documentclass[%
 reprint,
showpacs,preprintnumbers,
showkeys,
amsmath,amssymb,
aps,
]{revtex4-1}

\usepackage[utf8]{inputenc}
\usepackage[T1]{fontenc}

\usepackage[dvipsnames,table]{xcolor}

\usepackage{amsmath,amssymb}
\usepackage{nicefrac}
\usepackage[british]{babel}
\usepackage{booktabs}
\usepackage[justification=centering]{caption}
\usepackage{float}
\usepackage{csquotes}

\usepackage{tabularx}
\usepackage{graphicx}
\usepackage[export]{adjustbox} % align figures left, center, right
\usepackage{wrapfig} % wrap text around figures
\graphicspath{{./figures/}}
\usepackage{dcolumn}% Align table columns on decimal point
\usepackage{bm}% bold math
\PassOptionsToPackage{hyphens}{url}
\usepackage{hyperref}% add hypertext capabilities
\usepackage{listings}
\usepackage{multirow}
\usepackage{siunitx}
\DeclareSIUnit\feet{ft}

\usepackage{subcaption}
\usepackage{url}
\usepackage{enumitem}

\usepackage{ulem}

\usepackage[obeyFinal]{todonotes} % obeyDraft / obeyFinal: todonotes will be displayed only in draft / not displayed in final mode

\usepackage[pscoord]{eso-pic}
\newcommand{\placetextbox}[3]{% \placetextbox{<horizontal pos>}{<vertical pos>}{<stuff>}
  \setbox0=\hbox{#3}% Put <stuff> in a box
  \AddToShipoutPictureFG*{% Add <stuff> to current page foreground
    \put(\LenToUnit{#1\paperwidth},\LenToUnit{#2\paperheight}){\vtop{{\null}\makebox[0pt][c]{#3}}}%
  }%
}%

\newcommand{\ie}{that is} % this is better, when you read aloud, also better English
\newcommand{\eg}{e.\,g.}
\newcommand{\ea}{et al.~}
\newcommand{\code}[1]{\texttt{#1}}

\newcommand{\sref}[1]{Sec.\,\ref{#1}}
\newcommand{\fref}[1]{Fig.\,\ref{#1}}

\newcommand{\tref}[1]{Tab.\,\ref{#1}}

\newcommand{\sars}{\mbox{SARS-CoV-2}}

\newcommand{\SR}{S.\,R.}
\newcommand{\MG}{M.\,G.}
\newcommand{\GK}{G.\,K.}
\newcommand{\GH}{G.\,H.}

\setlist[description]{leftmargin=1em}

\begin{document}

\placetextbox{0.5}{0.99}{{\color{red}\rule{2\linewidth}{0.5mm}}}%
\placetextbox{0.5}{0.9725}{{\color{red}
The revised version of this preprint is now published in PLOS ONE (\href{https://doi.org/10.1371/journal.pone.0273820}{doi:10.1371/journal.pone.0273820}).}}%
\placetextbox{0.5}{0.96}{{\color{red}\rule{2\linewidth}{0.5mm}}}%

%\preprint{APS/123-QED}

%{\color{red} \rule{2\linewidth}{0.5mm}
%
%The revised version of this preprint is now published in PLOS ONE (\url{https://doi.org/10.1371/journal.pone.0273820).}
%
%\rule{2\linewidth}{0.5mm}}

\title{Modelling airborne transmission of \sars{} at a local scale}% Force line breaks with \\

\author{Simon Rahn}
\email{simon.rahn@hm.edu}
%\homepage{https://orcid.org/0000-0002-0842-2406}
\affiliation{Munich University of Applied Sciences HM, Department of Computer Science and Mathematics, 80335 Munich, Germany}
\affiliation{Technical University of Munich, Department of Informatics, 85748 Garching, Germany}

\author{Marion G\"odel}
\email{marion.goedel@hm.edu}
%\homepage{https://orcid.org/0000-0002-1572-842X}
\affiliation{Munich University of Applied Sciences HM, Department of Computer Science and Mathematics, 80335 Munich, Germany}
\affiliation{Technical University of Munich, Department of Informatics, 85748 Garching, Germany}

\author{Gerta K\"oster}%
\email{gerta.koester@hm.edu}
%\homepage{https://orcid.org/0000-0002-3369-6206}
\affiliation{%
 Munich University of Applied Sciences HM, Department of Computer Science and Mathematics, 80335 Munich, Germany
}

\author{Gesine Hofinger}%
\email{gesine.hofinger@team-hf.de}
% \homepage{}
\affiliation{%
 Team HF, Hofinger, K\"unzer \& M\"ahler PartG, 71634 Ludwigsburg, Germany
}

% \collaboration{MUSO Collaboration}%\noaffiliation

\date{17th November 2021}% Any date may be explicitly specified

\begin{abstract} % 198/200 words
The coronavirus disease (COVID-19) pandemic has changed our lives and still poses a challenge to science.
Numerous studies have contributed to a better understanding of the pandemic. In particular, inhalation of aerosolised pathogens has been identified as essential for transmission.
This information is crucial to slow the spread, but the individual likelihood of becoming infected in everyday situations remains uncertain.
Mathematical models help estimate such risks.
In this study, we propose how to model airborne transmission of \sars{} at a local scale.
In this regard, we combine microscopic crowd simulation with a new model for disease transmission. Inspired by compartmental models, we describe agents' health status as susceptible, exposed, infectious or recovered.
Infectious agents exhale pathogens bound to persistent aerosols, whereas susceptible agents absorb pathogens when moving through an aerosol cloud left by the infectious agent. The transmission depends on the pathogen load of the aerosol cloud, which changes over time.
%independent of the infectious agents' position.
%
We propose a \enquote{high risk} benchmark  scenario to distinguish critical from non-critical situations.
Simulating indoor situations show that the new model is suitable to evaluate the risk of exposure qualitatively and, thus, enables scientists or even decision-makers to better assess the spread of COVID-19 and similar diseases.
\end{abstract}

%\pacs{Valid PACS appear here}% PACS, the Physics and Astronomy Classification Scheme.
% three up to six keywords
% Use showkeys class option if keyword display desired

\maketitle

%\tableofcontents

\section{Introduction}
% General introduction
The outbreak of coronavirus disease (COVID-19) started at the end of 2019. Within months, it spread around the world and, ultimately, the World Health Organization~\cite{who-2020b-life} characterised COVID-19 as a pandemic on 11 March 2020.
%
% present perfect because it still affects us?
It has affected many aspects of our daily lives.
%
%Decision-makers and society as a whole have struggled with fighting the pandemic partly due to lack of knowledge about the disease.
%At the same time, scientists from various fields of research have tried to answer urgent questions about the coronavirus and how it spreads.
Therefore, scientists from various disciplines have attempted to answer urgent questions about the coronavirus and how it spreads.

% State-of-the-art
% Virology
% transmission via inhalation (airborne, droplet) (instead of indirect vs. direct)
From virology, we know that COVID-19 is caused by the severe acute respiratory syndrome coronavirus type $2$ (\sars{}). It is predominantly transmitted via respiratory fluids \cite{miller-2020-life, zhou-2021b-life}.
% 
% Distinction between smaller and larger droplets
This refers to both airborne and droplet transmission. However, there is no clear line between smaller, airborne droplets and larger ones \cite{gaef-2020b-life}.
%
% explain why we focus solely on airborne: z.B. weil Social distancing maßnahmen (nicht im Simulator) und Masken an vielen Stellen schon “in place” sind – die hauptsächlich bei Tröpfcheninfektion greifen
In this work, we focus on particles that remain airborne for at least a few minutes. We refer to these particles suspended in air \textit{aerosol clouds}.

% Epidemiology / Modeling community
% General intro
Modelling plays an important role in epidemiology.  Computer simulations help better understand the dynamics of a pandemic when ethical concerns prohibit experimental studies.
In the following, we discuss several approaches whose scope ranges from large to small populations.
%
%
%\begin{figure}[H]
%	\centering
%	\includegraphics[page=1, width=\linewidth]{./modelingApproaches/chart.pdf} 
%	%TODO ask MG about placement of "Wells Riley" bubble
%	\caption{Classification of our model for disease transmission among other approaches that: We distinguish different approaches according to their level of abstraction and the typical size of the problems they address. 
%	Our model belongs to the more detailed ones. It is agent-based and aims at simulating the transmission between individuals at a local scale over a short period.
%	In contrast, more abstract models usually capture the dynamics of a whole epidemic or pandemic.
%	} \label{fig:IntroModelingApproaches}
%\end{figure}

% [large scale] Compartmental SIR model
In their compartmental SIR model, Kermack and McKendrick~\cite{kermack-1927-life} describe the course of an epidemic, more precisely, the number of susceptible (S), infectious (I) and removed (R) individuals among a population, with ordinary differential equations.
%
% [large scale] modifications of the SIR model
This model was modified, \eg{} by adding an exposed state (SEIR) to simulate the latent period
% exposed state introduced by Anderson, R. M. & May, R. M. Infectious diseases of humans: dynamics and control. Oxford and New York: Oxford University Press. (1991). ?
or by accounting for possible reinfection after temporal immunity (SIRS).
The deterministic approach approximates a stochastic process of contact networks and, thus, is only valid for large populations.

% [large scale] network models
% - Newman, Spread of epidemic disease on networks, https://doi.org/10.1103/PhysRevE.66.016128
% - 
%Network models as applied in \cite{} represent another possibility for estimating the dynamics of a disease among a larger population. However, they are too abstract to gain information about the transmission of pathogens between individuals.

% [large/small scale] agent based models
In contrast, agent-based models capture the spread of diseases among smaller populations.
All virtual persons, the so-called agents, possess individual properties, such as their health condition. The transition from one status to another, \eg{} from susceptible to infected, depends on predefined rules.  
% examples:
% not yet calssified: kerr-2020-life, talekar-2020life
% larger:
% - abadeer-2020-cdyn: Vadere + disease spreading model (distance, ?probability?) , City of Muenster
% smaller: 
% - gosce-2014-cdyn: SFM + disease spreading model, corridor scenario
% - johansson-2013-cdyn: SFM + disease spreading model (distance, probability), mekka kaaba scenario
% - namilae-2020-cdyn: SFM + disease spreading model (small distance -> contact transmission, lager distance -> airborne transmission), airplane scenario for various diseases
% - ronchi-2020-cdyn: occupant exposure -> only evaluation of mutual distances / relationships / interactions
These rules are often based on mutual distances. For example, agents close to an infectious agent become infected after a certain time (see \cite{abadeer-2020-cdyn, gosce-2014-cdyn, johansson-2013-cdyn, namilae-2020-cdyn}).
Ronchi and Lovreglio \cite{ronchi-2020b-cdyn} expand the concept of proximity to further contact types, \eg{} physical contact, proximity within a certain radius and occupancy of the same room or building. 
The overall time spent in contact determines the risk of exposure.
Generally, agent-based models focus on contact time and proximity.
However, they neglect that aerosolised pathogens can remain at a position where they were emitted after the infectious agent has gone. Thus, infection is possible without obvious proximity.

% [small scale] CFD
Even if the infectious agent does not move, airflow may spread the pathogens and cause infections at distant places. Such transport mechanisms can be simulated with computational fluid dynamics (CFD) models.
% application / examples
For example, Vuorinen \ea\cite{vuorinen-2020-life} simulate how the aerosols of a coughing person travel. 
%The study shows that airborne transmission should be taken into account to get a better understanding of the transmission of \sars{}.
% advantage
%CFD simulations provide specific information about the spread of aerosols in the presence of airflow.
% disadvantage
However, CFD simulations require much, often uncertain information to set boundary conditions and, thus, to correctly capture a scenario. Even simple problems are computationally expensive. They become more demanding if CFD and crowd simulations are coupled, prohibiting scenarios with more than a few agents. 
Apart from that, the degree of detail between these two models does not match.
% consequence

% Motivation
% - covid-19: we need more information about local transmission
% - agent-based models / in particular microscopic crowd simulation is capable of modeling both infection and measures such as physical distancing
%
% Research gap:
% - existing agent-based models do not take into account "airborne transmission" (smaller aerosol particles), although this is a major mode of transmission
% - we need to answer the question: How can we model airborne transmission of SARS-CoV-2 adequately for everyday life situations?
% Motivation
In summary, macroscopic compartmental models, agent-based models and CFD simulations allow analysing the spread of diseases on different scales. Macroscopic models consider the overall dynamics of an epidemic, whereas microscopic models focus on pathogen transmission between individuals.
Gaining knowledge about the transmission on a local scale is of particular interest in the context of COVID-19.
Small-scale proximity models neglect that pathogens may persist in aerosols, even if their source is no longer close.
%
% research question
We wish to help bridge this gap and ask this question: How can we model airborne transmission of \sars{} between individuals for everyday situations?

% outline of the paper
To answer this research question, we first introduce the methodology and our microscopic crowd model
in \sref{sec:methods_and_materials}.   
% modeling
In \sref{sec:math_model}, we formulate a mathematical model for pathogen transmission via aerosol clouds.
We then computerise the model and couple it with crowd simulation in \sref{sec:comp_model}.
We calibrate the parameters in \sref{sec:calibration}  
to match \sars{}.
%We also build a reference scenario that allows us to translate the agents' degree of exposure back into the real world.
We also propose a reference scenario to which 
an agent's degree of exposure in any other situation can be compared.
Verification and validation in \sref{sec:verification_validation} show that the model is implemented correctly and that its results reflect empirical data.
% exemplary application
In \sref{sec:application}, we simulate a situation that is relevant for everyday life: transmissions between pedestrians waiting in a queue.
% conclusion
\sref{sec:conclusion_outlook} summarises and provides an outlook. % of this manuscript.

\section{Methods and materials}  \label{sec:methods_and_materials}
We adopt the classical modelling approach to build a new model for disease transmission. That is, we translate real-world observations into a mathematical formulation.
Then, we implement this model as an algorithm and generate a calibrated simulation programme.
This creates a virtual world in which we test various scenarios against empirical data.
Verification and validation are essential steps to ensure that the software contains very few errors 
% no software is error free
and that it yields results consistent with empirical observations. We validate our model by re-enacting two superspreading events. % and comparing the results to the original records.

The transmission model is integrated into Vadere \cite{kleinmeier-2019-cdyn}, an established framework for microscopic crowd simulation. We use the new sub-model in combination with the state-of-the-art Optimal Steps Model (OSM) \cite{seitz-2012-cdyn, sivers-2015-cdyn}, but it is also compatible with other locomotion models.
%
%Vadere including the new transmission model is implemented in Java.
%
%For data curation, we use Python 3.
%
The source code and all relevant data are publicly available on Gitlab \cite{vadere-2021-cdyn}.

\section{Mathematical model for transmission via inhalation}  \label{sec:math_model}
% Transmission via inhalation is the main path of infection for \sars{} (see Absorption of pathogen)
In this section, we model the transmission of \sars{} among individuals via exhalation and inhalation. %, since the virus spreads mainly through pathogens bound to aerosol particles. 
%Direct infection seems to play a minor role. For this reason, we focus only on transmission via inhalation.
%
We develop our model specifically for COVID-19 but it can be transferred to other diseases that also spread through pathogens bound to aerosol particles, such as influenza \cite{tellier-2009-life}.
% Observations: Literature / meta studies on disease transmission modes (transmission via aerosol clouds, properties of aerosol clouds and pathogen particles in aerosol clouds, etc.) 
% Description of mathematical model
As a first step, we operationalise real-world observations and derive a mathematical model.

\subsection{The agents' state of health}
%\gkcomment{Was ich unnötig kompliziert und auf den Leser fehlleitend finde, ist die Auseinandersetzung mit "recovered" Agents und den Übergang zu infectious. Das spielt auf unserer Zeitskala beides einfach gar keine Rolle. Ich schlage vor, das auch so zu sagen:
%On our time scale of minutes to an hour, recovery and the transit from exposed or infected to infectious play no role.
%Und dann lässt man auch die genauen Definitionen weg.
%Vorteil: Es ist von Anfang an klar, was wir nicht machen.}
%
%\srcomment{Stimmt, die Übergänge E $\rightarrow$ I, I $\rightarrow$ R, R $\rightarrow$ S sind von der Zeitskala her nicht relevant und nur der Vollständigkeit halber für den Fall implementiert, dass jemand wie Mina Abadeer längere Zeiträume simulieren will. Ich habe versucht das umzuformulieren. Die Übergänge habe ich aber bisher nur teilweise rausgenommen, da m.E. der Unterschied zu dem ursprünglichen SEIR Modell klar werden sollte. Ist das nun besser verständlich, oder Übergänge ganz raus?}
%
%\srcomment{Der Status R ist im Modell m.E. relevant, um z.B. immunisierte Personengruppen abzubilden.}

% Infection status (SEIR) & transition rules
%TODO Check anderson-1992-life -> Introduces status E?
Inspired by the compartmental SIR model \cite{kermack-1927-life}, we define an agent's health as  susceptible (S), exposed (E), infectious (I) and recovered (R).
\textbf{Susceptible} agents represent healthy persons. They inhale pathogens and accumulate them. Once the minimum infectious dose is exceeded, their status changes to exposed.
\textbf{Exposed} agents are considered infected. They retain the attributes of susceptible agents but will become infectious after a latent period. This transition plays no role in our contribution since the time scale of our simulations ranges from minutes to a few hours, which is significantly shorter than the latent period for \sars{}.
%
%During the period of communicability, we consider all agents \textbf{infectious}. 
\textbf{Infectious} agents emit pathogens via aerosols. For simplicity, we do not distinguish between symptomatic and asymptomatic cases and keep the infectiousness constant.
\textbf{Recovered} agents do not absorb pathogens, \ie{} they are immune. Since the simulation time is short, they occur only if they are included in the initial condition of the simulation.

\begin{figure}[h]
	\centering
	\includegraphics[width=\linewidth]{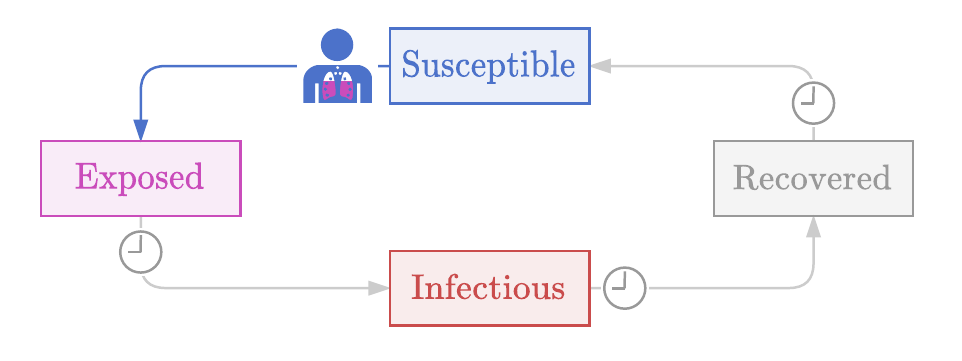} 
	\caption{
	An agent's health states: The transition from susceptible to exposed is relevant for the
	time scale of our model.
		} \label{fig:MathModelInfectionStatusCycle}
\end{figure}

% Transition order
The cycle in \fref{fig:MathModelInfectionStatusCycle} 
visualises the transition between health states, focussing on the transition from susceptibility to exposure.

%Besides, we neglect time-dependent effects such as a varying infectiousness. 
%
% Update of health status – e.g. absorbing pathogen load
Independent of the current infection status, each agent has a respiratory cycle of equally long periods of inhalation and exhalation. Pauses in between are neglected. Hence, we obtain the respiratory frequency $f$ and the corresponding period $T=f^{-1}$.
During exhalation, infectious agents emit pathogen bound to aerosol clouds. Susceptible agents inhale a fraction of these pathogens if their current position is within the aerosol cloud. 
%The fraction is determined by the volume inspired and expired with each breath the tidal volume, \ie{} , determined by the tidal volume, of the pathogen concentration that is present at their current position.
%This concentration depends on the number of particles that have been emitted by an infectious agent at an earlier point in time.

\subsection{Emission of pathogen}
% Intro
In the case of COVID-19, infectious persons emit pathogen mainly through aqueous droplets expelled during breathing, but also, \eg{} by speaking or coughing.
These expiratory events vary in intensity and, thus, in droplet numbers and droplet sizes \cite{gaef-2020b-life}, which in turn alters how
the droplets spread through the air.
Violent expiration causes larger droplets, which follow a ballistic trajectory, whereas normal breathing produces smaller droplets \cite{gaef-2020b-life} that remain in the air and form a cloud around the source. 
%
% definition of aerosol
In either case, there is a suspension of liquid particles in the air, which is often called aerosol.
There is no clear line between small and large particles \cite{uba-2021-life}. In this contribution, we consider particles that stay airborne for at least a few minutes.
% creation
At this point, we focus on normal breathing. However, the concepts we present can be transferred to any kind of aerosol producing activity as long as the particles are small enough to stay airborne for some minutes.

\begin{figure}[h]
    \begin{subfigure}{0.45\linewidth}
        \includegraphics[width=\linewidth]{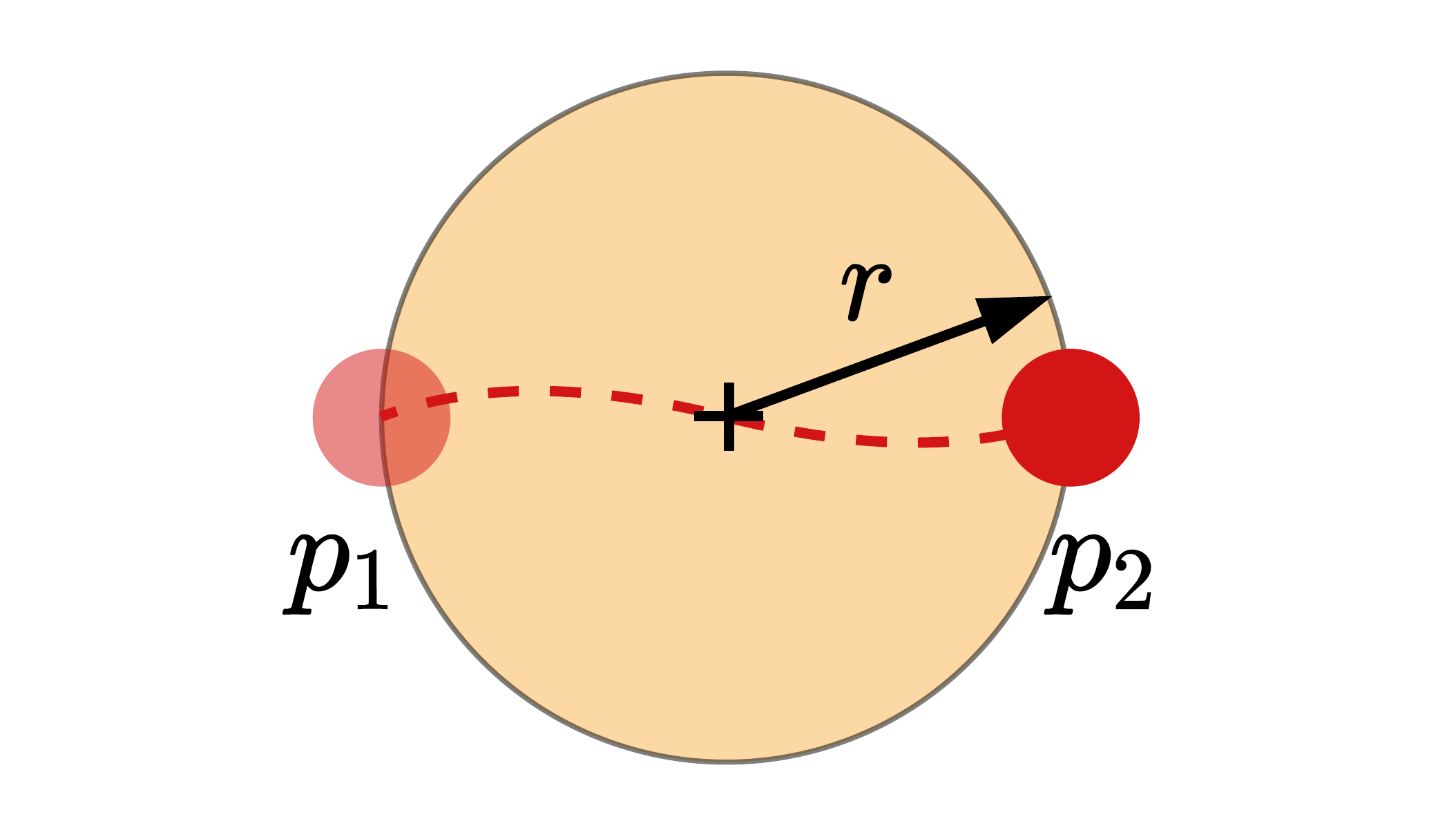}
        \caption{} \label{fig:MathModelTransmissionShapesAerosolCloudCircle}
    \end{subfigure}
    \begin{subfigure}{0.45\linewidth}
        \includegraphics[width=\linewidth]{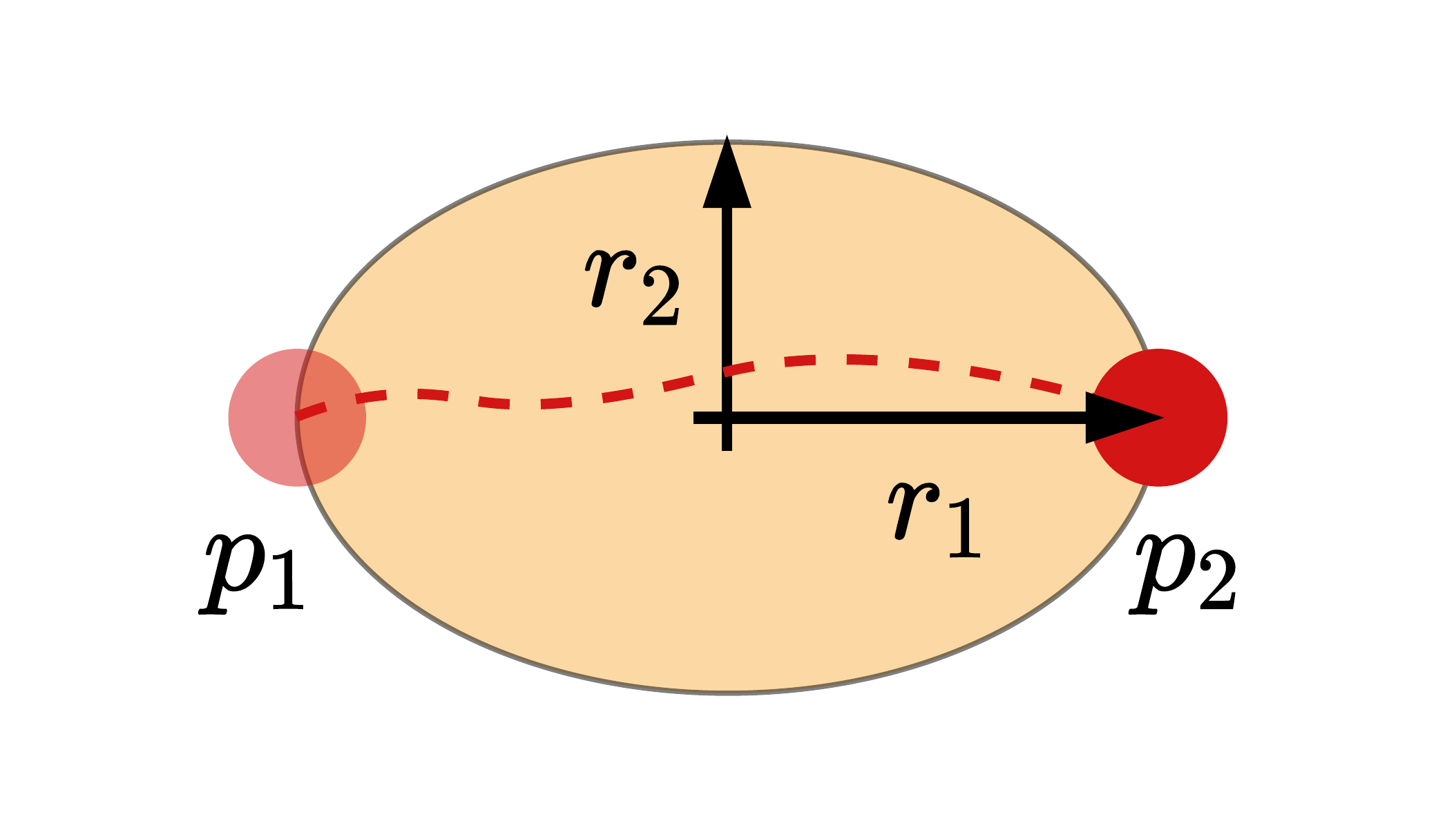}
        \caption{} \label{fig:MathModelTransmissionShapesAerosolCloudEllipse}
    \end{subfigure}
    \caption{An infectious agent (red) emits aerosol clouds (orange). The shape of the cloud depends on the agent's speed. Low velocities lead to circular clouds (a) with radius $r$, whereas higher velocities yield elliptical clouds (b) with semi-axes $r_1$ and $r_2$. The dashed line represents the agent's trajectory from left to right. The agent starts exhaling at $p_1$ and stops  at $p_2$.}
    \label{fig:MathModelTransmissionShapesAerosolCloud}
\end{figure}

An infectious agent creates an aerosol cloud containing pathogen particles with every exhalation.
See \fref{fig:MathModelTransmissionShapesAerosolCloud}.
%
%We neglect aerosols that are emitted by non-infectious individuals.

% geometry / shape
We are aware that particle clouds can be very complex. However, we argue that modelling such detail would be unsuitable for the degree of abstraction of the overall model. Besides, airflow is often turbulent, making its computation a major and still unsolved problem. Hence, we decided to keep the model simple:
since pedestrian dynamics are typically modelled in two dimensions, we represent each aerosol cloud by a two-dimensional shape in the horizontal layer. % We imagine it at the height of the agent's head.
The initial cross-section $A_0$ is equal for all clouds.
In the case where the agent moves slowly, the shape is a circle with a radius $r$ (\fref{fig:MathModelTransmissionShapesAerosolCloudCircle}).
We let the cloud form over the entire period of exhalation $T$, so the shape is centred between the position, where the agent starts breathing out $p_1$, and the position, where it stops $p_2$.
In a case where the agent walks faster, \mbox{$\left| v \right| > 2\, r T^{-1}$}, we obtain an ellipse with vertices at $p_1$ and $p_2$ (\fref{fig:MathModelTransmissionShapesAerosolCloudEllipse}), with the cloud growing longer as the agent walks faster.

% "evolution" of the extent of the cloud
% optional: list scenarios from the case study (examples for scenarios that can be modelled w/o airflow)
%
The extent of aerosol clouds changes over time, in particular, in the presence of airflow.
Again, we choose to focus on the simplest situation, with stagnant air, as can be found in confined spaces without ventilation. Such places are deemed particularly risky for \sars{} transmission. %More complex flow situations should be considered in the future.
Any agent $i$ who walks through a cloud with velocity $v_i$ disperses aerosol particles. To capture this scenario, we enlarge the clouds' radius by
$\Delta r \left( t \right) =  \Delta t \sum_{i} \left| v_i \left( t \right) \right|$
at each time step $t$. Factor $\Delta t$ is the length of a simulation step. It ensures that the dispersion model is independent of the user's choice for  $\Delta t$. 
%Weight $w$ allows the user to adjust the influence of the agents' walking speeds on the dispersion.
In the case of an elliptical cloud, we proportionally increase the semi-axes $r_1$ and $r_2$. 

% distribution of pathogens in 3D
In reality, aerosol clouds spread in three dimensions. In addition, we need a model of pathogen concentration in a volume that agents can breathe in.
Since we are unaware of empirical evidence on the spread of the aerosolised pathogen in a stagnant air volume, we stick to a simple description: a homogeneous distribution.
We represent the aerosol clouds with two-dimensional shapes in a horizontal layer at the height of the agents' heads. From there, we imagine them to extend to the same volume as a sphere with a radius $r$: $V=\frac{4}{3} r^3 \pi$.  
%
% a)
%Due to lack of knowledge, we assume a homogeneous distribution of pathogens within the cloud and  spherical spreading yields the volume $V = r^3 \, \pi$. Since we consider only the horizontal layer at the agents' heads, we do not distinguish between elliptical and circular cross-sections. It is sufficient to account for a vertical extent $h = \frac{V}{A_0}$ so that the cloud's initial volume equals the volume of a sphere with radius $r$.
%
% b)
%
Thus, we can measure the clouds' pathogen load as concentration $C_p$ in particles per cubic metre, which decreases when the cloud extends.
Furthermore, we assume that the concentration accumulates where clouds overlap.
% exponential decay
%Also the temporally declining number of pathogens within an aerosol cloud yields a decreasing concentration. Pathogens can be inactivated, so they no longer contaminate the aerosol. The pathogen carrying aerosol particles can be removed from the considered layer. They evaporate, sediment due to gravity, or rise to higher air layers. To map all these effects, we assume an exponential decay of the number of pathogens with a half-life $T_a$.
The initial pathogen load is equal for all aerosol clouds that are emitted by the same agent. This initial number of pathogens depends on the individual's infectiousness.
The pathogens of an aerosol cloud are inactivated after some time. In addition, aerosol particles evaporate, rise to higher air levels, or sediment.
We simplify these complex processes by assuming an exponential decay of the number of pathogens with a half-life $T_a$.
%
%Furthermore, the number of pathogens contained in an aerosol cloud decreases over time. This depends on both the pathogen itself and the behavior of the aerosol particles. Pathogens can be inactivated, so they no longer contaminate the aerosol. The pathogen carrying aerosol particles can be removed from the considered layer. They can evaporate, sediment due to gravity, or rise to higher air layers. To map all these effects, we assume an exponential decay of the number of pathogens with a half-life $T_a$.
%
%As soon as the aerosol cloud reaches less than $1\%$ of its initial pathogen concentration, we consider it diluted and therefore negligible.

\subsection{Absorption of pathogen}
% main path of pathogen transmission: inhalation
%susceptible agents inhale pathogens, once they are present in an aerosol cloud.
%
%The model does not include other possible ways, \eg{} having infectious droplets land directly or being transported indirectly on ocular, oral, nasal, or other mucous membranes.
%
% absorbtion
While in the vicinity of one or more aerosol clouds, susceptible and exposed agents breathe in pathogens. Each agent absorbs the number of pathogen particles $N_p$ contained in the tidal volume $V_T$. The tidal volume is the volume inhaled and exhaled with each breath \cite{lufti-2017-life}. From this follows $N_p = C_p \, V_T \, \left( 1 - E_p \right)$. Masks reduce $N_p$ by their effectiveness $E_p$. $C_p$ is the sum of the pathogen concentrations of all aerosol clouds at the agents' position. 
We neglect that inhalation removes pathogens from the surroundings.

% accumulation
The number of pathogens accumulated by an agent describes its degree of exposure.
%
%If the minimum infectious dose for \sars{} is known, one can also decide whether the agent becomes actually infected or not.
%
The minimum infectious dose is the number of pathogens required for infection. In the model, it marks the transition from susceptible to exposed. Its calibration poses a challenge since, at present, there is no consensus on the right value for \sars{}. We will discuss this further in \sref{sec:calibration}.

% Eventuell:
% Assumptions , considered / neglected aspects – areas of application / scenarios

\section{Computerised model}  \label{sec:comp_model}
% Introduction
%The mathematical model aims to describe the transmission of pathogens through aerosol clouds.
%
Here, we combine the transmission model with the OSM \cite{seitz-2012-cdyn, sivers-2015-cdyn}, a state-of-the-art locomotion model for pedestrian dynamics 
implemented in the open-source framework Vadere \cite{kleinmeier-2019-cdyn}.
%
%The locomotion model represents pre-Covid conditions and the transmission model does not affect the locomotion model. In reality, the pedestrians' walking behavior could be affected by reservations towards apparently ill persons and result in keeping a distance or avoiding crowded areas. However, one can adapt the locomotion model as in \cite{mayr-2020-cdyn} in order to account for physical distancing measures or other restrictions.
%
Vadere is well-established for crowd simulations, which is why we adhere to its software architecture when we define important requirements for a new feature:
%
% Goals
The code should be compatible with existing structures,
modular so that the transmission model can easily be enabled or disabled for different locomotion models and flexible to allow for adaptations by other developers.

\subsection{Embedding in Vadere}
This section covers the embedding of the transmission model in Vadere. It gives insight into the software structure and how developers can enhance the model or implement additional features. 
%\sref{sec:Parameters} shows users how to adapt the model according to their needs. % Fand den Satz hier eher verwirrend. Sehr feinsinnige Unterscheidung zwischen enhancing und adaptation mit und ohne Programmieren

\begin{figure}[H]
	\centering
	\includegraphics[width=\linewidth]{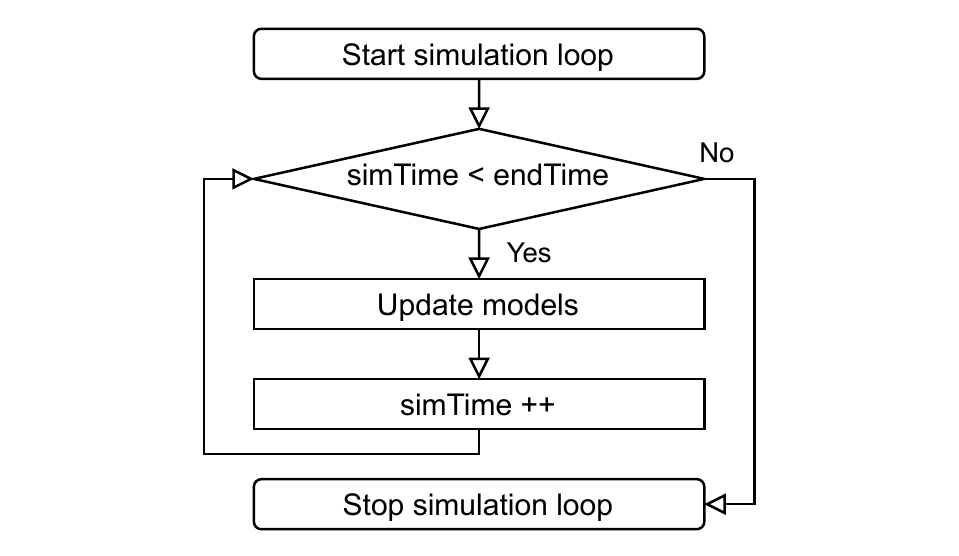}	
	\caption{Models are updated %for each simulation step 
	as long as the simulation loop is running.} \label{fig:CompModelUMLflowChart}
\end{figure}

\begin{figure}[h]
	\centering
	\includegraphics[width=\linewidth]{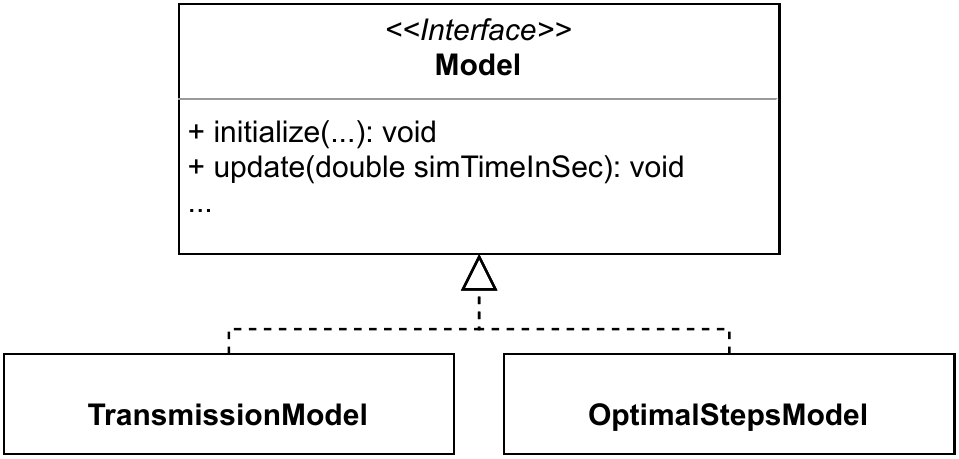}
	\caption{\code{TransmissionModel} contains the model core. Similar to the locomotion model \code{OptimalStepsModel}, it implements the interface \code{Model}.} \label{fig:CompModelUMLmodel}
\end{figure}

% Flow
The transmission model is integrated into Vadere's simulation loop, a while-loop that updates all elements, mainly sources, targets and agents' positions, as long as the simulation is running (\fref{fig:CompModelUMLflowChart}). 
The simulation loop calls the transmission model independently of other models, keeping the programme modular and flexible. In particular, this allows combining transmission with any Vadere's locomotion model.
Note that a user must select a locomotion model when running Vadere, whereas using the transmission model is optional.
% Structure / Classes of the transmission via inhalation model
\fref{fig:CompModelUMLmodel} visualises the structural embedding of the transmission model:  
The \code{TransmissionModel} implements the interface \code{Model}, as the locomotion models in Vadere do.
Supplementary features or alternative models for disease transmission can be added on the same level.

% Attributes of transmission model / Model parameters (input file)
\code{TransmissionModel}'s attributes contain general model parameters. Developers can adapt these attributes in the class \code{AttributesTransmissionModel}. We describe the meaning of each parameter in detail in \tref{tab:CompModelParameters}.

% Methods / logic
The methods of the class \code{TransmissionModel} contain most of the model's logic. In essence, the method \code{update(\dots)} controls the emission of the pathogen and updates aerosol clouds and the agents' health status.
Each time an infectious agent stops exhaling, an aerosol cloud is inserted into the topography. Initially, the aerosol clouds cover the distance passed during the exhalation but can expand over time, whereas their pathogen load decreases exponentially. Aerosol clouds at the end of their lifetime are removed.
Furthermore, the update method induces susceptible and exposed agents to absorb pathogens if they are in aerosol clouds.
The routine also updates each agent's respiratory cycle and infection status.

\begin{figure}[h]
	\centering
	\includegraphics[width=\linewidth]{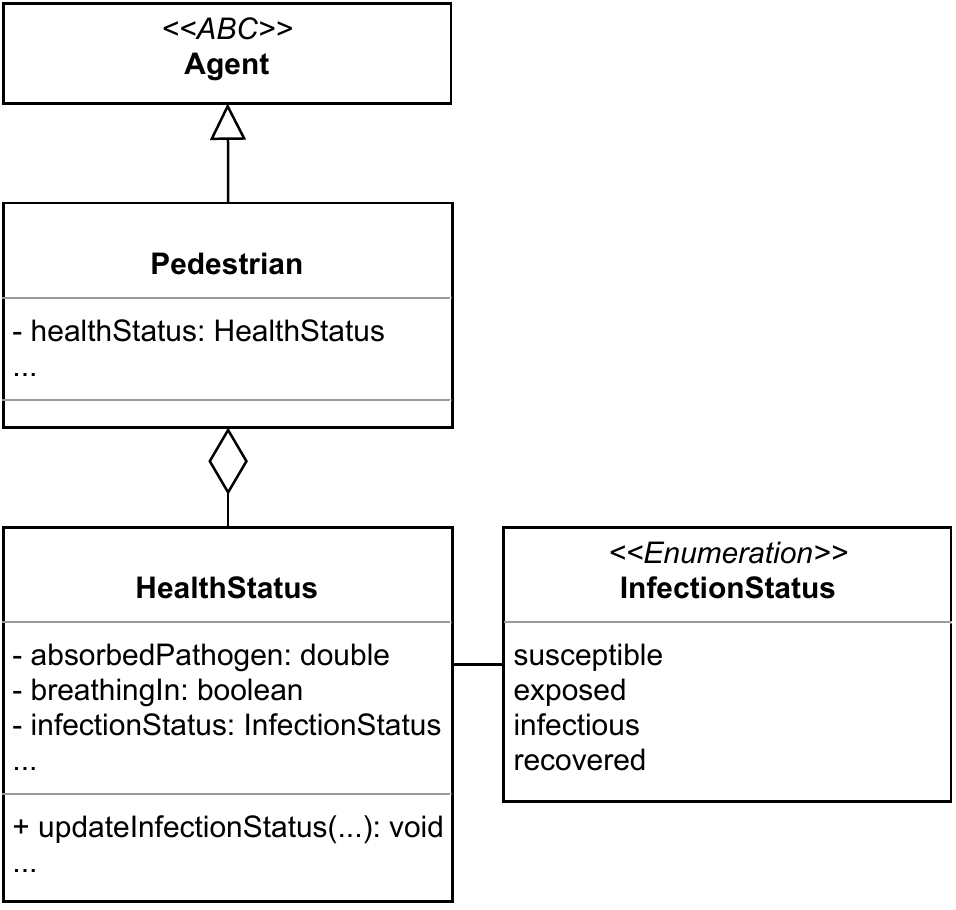}
	\caption{Pedestrians' attributes wrapped in the class \code{HealthStatus}.} \label{fig:CompModelUMLhealthStatus}
\end{figure}

% Health status
An agent's health status is wrapped in the class \code{HealthStatus}, which is an attribute in \code{Pedestrian} as shown in \fref{fig:CompModelUMLhealthStatus}. \code{HealthStatus} contains absorbed pathogen load, respiratory cycle and infection status
as its most important attributes.

% Aerosol clouds
Aerosol clouds are embedded as \code{InfectiousParticleDispersion} that extends the class \code{ScenarioElement} and, thus, fit into the existing structure. Typical scenario elements are sources, targets or obstacles. This approach allows individual access and manipulation throughout the simulation runtime and facilitates the graphical representation of clouds.

\subsection{Model parameters} \label{sec:Parameters}
% Generelles einführen der Parameter in der Modellbeschreibung (weiter oben)
In this section, we summarise all important model parameters. Users can adapt the values directly in the input file.
The parameters apply to all agents and aerosol clouds within the same simulation. When an agent is spawned, the parameter values are assigned to the agent's attributes.

% parameters related to / used by agents health status
The agent's health and aerosol clouds are partly defined by attributes listed in \tref{tab:CompModelParameters}. 
We are aware that, in reality, some parameters related to the agents' health are time-dependent or differ from person to person. 
However, we do not expect significant changes within the simulation time, which is very short compared to the period of communicability. 
We capture the heterogeneity among infectious agents, \eg{} pathogen load or position, by running separate simulations for adapted parameter sets. %We do this because the information is  unavailable.

\section{Calibration}  \label{sec:calibration}
The model presented here allows simulating any disease transmitted by aerosols. We select a set of parameters to match the transmission of \sars{} through aerosol clouds formed by normal breathing.

% Default values
\subsection{Parameter set for \sars{}} \label{sec:calibrationSARS-CoV-2ParameterSet}
The parameters are physical in the sense that we can directly transfer them from the real into the virtual world. Their values are summarised in \tref{tab:CompModelParameters}. 

\begin{table*}[t]
\caption{Model parameters describing agents' health status and aerosol clouds. The values correspond to a highly infectious agent that exhales SARS-CoV-2.}
\label{tab:CompModelParameters}
\begin{tabularx}{\textwidth}{l l l X l l}
\hline
& \textbf{Parameter} & \textbf{Symbol} & \textbf{Description} & \textbf{Value} & \textbf{Unit} \\ \hline \hline
\multirow{7}{*}{\rotatebox[origin=c]{90}{\parbox[c]{4cm}{Agents' health status}}}
%
% pedestrianRespiratoryCyclePeriod
& \code{respiratoryCyclePeriod} & $T$ & Duration of one breath (inhalation and exhalation) & $4$ & $\si{\s}$ \\ \cline{2-6}
%
% pedestrianPathogenEmissionCapacity
& \code{pathogenEmissionCapacity} & $N$ & Number of pathogen particles emitted per exhalation by an infectious agent; logarithmised to base $10$ & $4$ & $\text{particles}$ \\ \cline{2-6}
%
% pedestrianPathogenAbsorptionRate
& \code{pathogenAbsorptionRate} & $R$ & Relation of absorbed pathogens to the pathogens that are present in the surrounding; Interpretable as the agents' average tidal volume $V_T$, potentially reduced by effectiveness $E_p$ of protective devices: \mbox{$R = V_T \, \left( 1 - E_p \right)$}
& $5 \cdot 10^{-4}$ & $\frac{\si{\cubic\m}}{\text{inhalation}}$ \\ \cline{2-6}
%
% pedestrianMinInfectiousDose
& \code{minInfectiousDose} & $D$ & Number of pathogens required for an infection, \ie{}, the threshold above which an agent changes from susceptible to exposed & $3.2 \cdot 10^3$ & $\text{particles}$ \\ % \cline{2-6}
%%
% exposedPeriod
%& \code{exposedPeriod} & $T_E$ & \multirow{3}{\hsize}{Periods defining the time-dependent transitions between the infections status E $\rightarrow$ I, I $\rightarrow$ R, and R  $\rightarrow$ S, respectively; relevant for longer simulations} & $2.59 \cdot 10^5$ & $\si{\s}$ \\
%
%& \code{infectiousPeriod} & $T_I$ & & $3.46 \cdot 10^5$ & $\si{\s}$ \\
%
%& \code{recoveredPeriod} & $T_R$ & & $1.56 \cdot 10^7$ & $\si{\s}$ \\
\hline
\multirow{2}{*}{\rotatebox[origin=c]{90}{\parbox[c]{1.6cm}{Aerosol cloud}}}
%
% aerosolCloudInitialArea
& \code{initialArea} & $A_0$ & The cloud's extent at creation time; circular with radius $r$ and elliptical  with semi-axes $r_1$ and $r_2$; \mbox{$A_0 = r^2 \, \pi = r_1 \, r_2 \, \pi$} & $7.1$ & $\si{\m\squared}$ \\ \cline{2-6}
%
% aerosolCloudHalfLife
& \code{halfLife} & $T_a$ & Defines the exponential decay of the pathogens in an aerosol cloud & $600$ & $\si{\s}$ \\ \hline
\end{tabularx}
\end{table*}

% breathing rate / respiratory cycle period
An adult at rest breathes approximately $12$ to $18$ times per minute. We use an average of $15$ breaths per minute, which implies a period of $T = \SI{4}{\s}$ between inhalations. The exact values depend on factors such as the level of physical activity. However, slightly different breathing rates affect the quantity of absorbed pathogens significantly less than other parameters. 
%
% emission capacity
The emission capacity in particular, which describes the number of emitted pathogens per breath, may vary in orders of magnitude for \sars{}.
Ma \ea\cite{ma-2020-life} found that some individuals exhale up to $400,000$ viral particles per minute. With $15$ breaths per minute, this means more than $10^4$ viruses per exhalation. Note that it is unclear if the pathogens were also emitted through larger droplets.
%
%In contrast to the high viral loads reported in \cite{ma-2020-life}, 
COVID-19 positive cases may emit significantly fewer pathogens, \eg{} when viral replication is low for their variant \cite{mlcochova-2021-life} or because they are not at the peak of their infectiousness.
Since we are interested in observing infection spread, we simulate a highly infectious person with $10^4$ particles per exhalation. To our knowledge, this represents the upper limit of a realistic range.

% absorptionRate
We now turn our attention to the absorbing, susceptible agents. The absorption rate $R$ can be interpreted as the tidal volume in cubic metres. For an adult, it is approximately $0.5$ litre, or $R = 5 \cdot 10^{-4} \,\si{\cubic\m}$, per inhalation. Masks would reduce this rate depending on their effectiveness.

% susceptibility / minimum infectious dose
The infectious dose $D$ of \sars{} in humans is still uncertain. It probably depends on the individual and the variant of the virus.
Karimzadeh \ea\cite{karimzadeh-2021-life} estimate that approximately $10^2$ particles cause an infection.
Popa \ea\cite{popa-2020-life} analyse epidemiological clusters and infer that, on average, more than $10^3$ viral particles can successfully start an infection, but smaller quantities may suffice.
Since this parameter is uncertain, we re-examine it in section \sref{sec:calibrationInfectiousDose}, where we calibrate the infectious dose through a reference scenario and set $D = 3200$.

% periods
%TODO check primary references for latent period, and half-maximum infectious period
%\gkcomment{For our investigation $T_E$, $TI$ and $T_R$ play no role. They only confuse the reader. I dropped this paragraph}
%
%The periods for how long an agent remains exposed ($T_E$), infectious ($T_I$), and recovered ($T_R$) are important if the simulation covers a period of several days to weeks:
%
%The exposed period is equivalent to the median latent period, which is approximately three days \cite{bar-on-2020-life}.
%
%The exposed period is followed by an infectious period of about four days. Transmission of pathogens is most likely during this time \cite{bar-on-2020-life}.
%We set the infectious period to $T_I = 4$ days accordingly
% Thereafter, we consider the agents recovered and immune.
%
%Wang \ea\cite{wang-2021c-life} report a time span of six to twelve months during which the antibody reactivity is stable. We set six months as recovered period.
% Connecting sentence
%In addition to the health status related parameters, we define the spread of \sars{} in the air through aerosol clouds.

The aerosol clouds are characterised by the following parameters: the initial area and the half-life.
% aerosol clouds
%The aerosol clouds are characterized by the initial area and the half-life.
%
We assume an initial radius of $r = \SI{1.5}{\m}$ for circular aerosol clouds and, thus, an area of approximately $A_0 = \SI{7.1}{\square\m}$. With the spheric extent described in \sref{sec:math_model}, we obtain a volume of $V = \SI{9.4}{\cubic\m}$. Hence, the initial pathogen concentration of an aerosol cloud is about $10^3$ particles per cubic metre.
%The initial area is based on the physical distancing measures taken during the COVID-19 pandemic. The prescribed mutual distances vary from country to country, often between $\SI{1}{\m}$ and $\SI{2}{\m}$. Therefore, we assume a radius of $r = \SI{1.5}{\m}$ for circular aerosol clouds, which yields an initial area of $A_0 = \SI{7.1}{\square\m}$.
%
%judgement based on 
We rely on reports from experience for the half-life of an aerosol cloud's \sars{} load.  
The half-life of artificially generated aerosols was found to last from $\SI{30}{\minute}$ to several hours 
\cite{doremalen-2020-life, smither-2020-life}. 
%These studies focus on the stability of \sars{} in artificially generated aerosols. However, aerosols emitted by humans are heterogeneous in size. Therefore, they partly remain shorter in the air than the virus is active. The pathogens sediment together with the aerosol particles. Further research is necessary to better estimate the distribution of aerosol particle sizes, how long they persist, and how long airborne pathogens remain infectious \cite{gaef-2020b-life}. For now, we estimate that the pathogen load decreases to half its original amount after %a half-life of ten minutes.
%
%TODO cite{smith-2020-life} ? half-life is only 50% if relative humidity is 100% -> no evaporation, aerosol particles fall faster, half live 5.5 ... 7 min
%
The exact value for the half-life is not so important if the model output is interpreted qualitatively, as is the case in this contribution.
However, it affects the dynamics of the model. A shorter half-life closely links exposure to the current location of an infectious agent. On the other hand, a long half-life means that agents can become exposed even if the infectious agent has left the area long ago.
%Although this parameter is physical, it is not easy to determine. A parameter study in the next section allows us to get a better understanding of its influence on the simulation output.

\subsection{Calibration of the infectious dose} \label{sec:calibrationInfectiousDose}
% Intro
We gained parameter values on the transmission of \sars{} from studies that have limitations.
%
%Since not all the necessary information is available to clearly set the parameters, 
Therefore, we propose to evaluate simulation scenarios in relation to a reference: the so-called \textit{close contact scenario}.
We choose the number of pathogens that a susceptible agent absorbs in this situation as our infectious dose $D$.

% close contact scenario
Governmental and institutional instructions regarding contact tracing during the COVID-19 pandemic define close contacts as follows:
A susceptible individual and a confirmed case of COVID-19 occupy an enclosed space without adequate ventilation. They are in close proximity for a certain time so that the susceptible person inhales aerosolised \sars{} particles. The leading scientific institute in Germany in the context of the COVID-19 pandemic, Robert Koch Institute~\cite{rki-2021c-life}, declares a distance of less than $\SI{1.5}{\m}$ for more than $\SI{10}{\minute}$ as critical. The Centers for Disease Control and Prevention, U.S. Department of Health and Human Services~\cite{cdc-2021-life}, specifies $\SI{6}{\feet} \approx \SI{1.8}{\m}$ for more than $\SI{15}{\minute}$.

\begin{figure}[ht]
\centering
\includegraphics[width=0.25\linewidth]{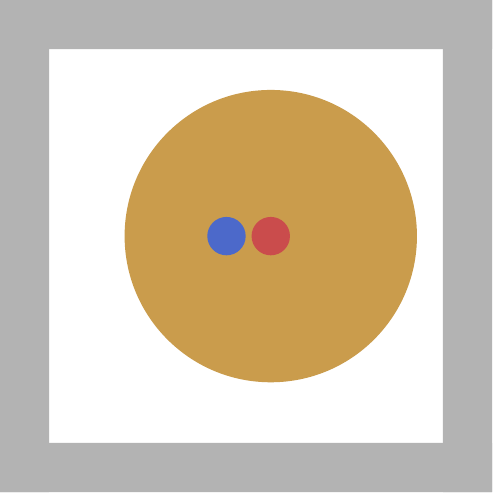}
\caption{\textit{Close contact scenario}: a highly infectious agent (red) emits pathogens bound to aerosols (orange) in an unventilated enclosed space. A susceptible agent (blue) absorbs pathogens.} \label{fig:CalibrationCloseContactScenarioVisualization}
\end{figure}

We follow the former definition with the parameter set from \tref{tab:CompModelParameters}. Two agents are placed less than $\SI{1.5}{\m}$ apart (\fref{fig:CalibrationCloseContactScenarioVisualization}). Both remain stationary for $\SI{10}{\minute}$.

\begin{figure}[ht]
\centering
\begin{subfigure}[b]{0.9\linewidth}
	\centering
	\includegraphics[width=\textwidth]{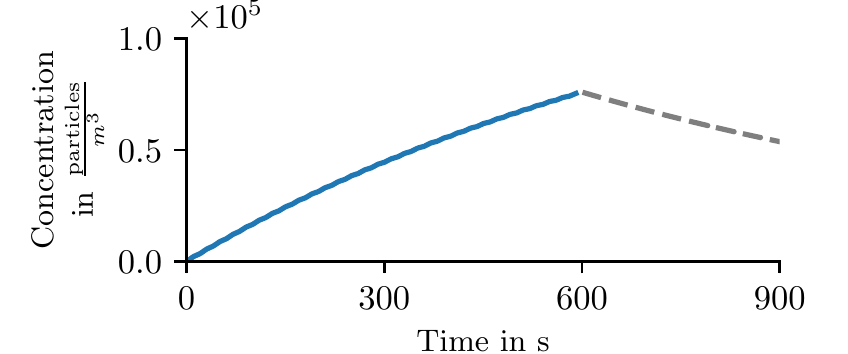}
    \caption{Pathogen concentration of aerosol clouds adds up.}
\end{subfigure}
\\
\begin{subfigure}[b]{0.9\linewidth}
	\centering
	\includegraphics[width=\textwidth]{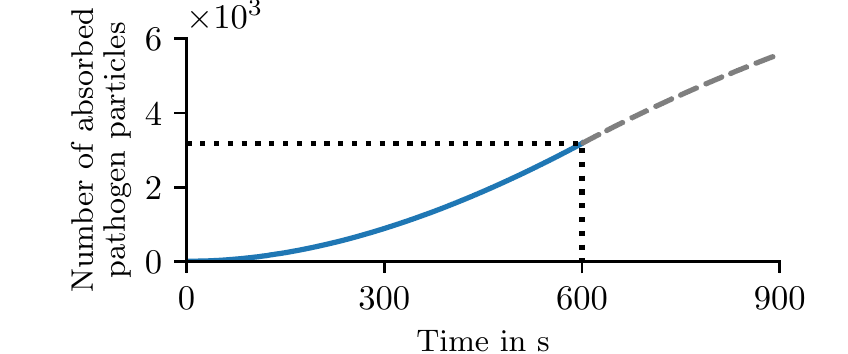}
    \caption{Pathogen load absorbed by the susceptible agent.}
\end{subfigure}
\caption{
\textit{Close contact scenario}: an infectious agent emits aerosol clouds so that the pathogen concentration builds up. The susceptible agent absorbs approximately $3200$ particles within the critical period of $\SI{600}{\second}$ (dotted lines).
The dashed lines represent an extension of the scenario: The infectious agent leaves at $t = \SI{600}{\second}$, whereas the susceptible agent remains and keeps absorbing pathogen from the exponentially decaying aerosol clouds.
} \label{fig:CalibrationCloseContactScenarioOutput}
\end{figure}

The infectious agent constantly emits aerosol clouds, thereby increasing the pathogen concentration. The susceptible agent absorbs approximately $3200$ pathogen particles within $\SI{10}{\minute}$ (see \fref{fig:CalibrationCloseContactScenarioOutput}).
%
%By means of the calibrated infectious dose, we can evaluate other scenarios in relation to the close contact scenario.
In all further simulations, agents who inhale $D = 3200$ or more pathogens are considered exposed.
% excursus: decay and absorption of pathogen in absence of an 

As soon as the infectious agent leaves the scenario, \eg{} at $t = \SI{600}{\s}$, the pathogen concentration decreases exponentially. If the susceptible agent remains, it will keep inhaling pathogen particles from the persistent aerosol clouds (see \fref{fig:CalibrationCloseContactScenarioOutput}).

% (All relevant discretization of the model: e.g. approximation of ellipses with rectangles …) 

\section{Verification and validation}  \label{sec:verification_validation}
% Intro
Generally, careful verification and validation are necessary parts of the modelling and software development processes.
%To assure the quality of the simulation program, we verify and validate our model.
%Vadere already provides test strategies that we apply to the new feature as well.
%
% Verification
We verify the transmission model by running unit tests.
The test coverage of the core of our model reaches approximately $80 \%$.
Vadere's continuous integration pipeline also contributes additionally to detecting errors in the code, as every commit is tested automatically.
%
% visual
%
The locomotion models have been verified with unit tests and validated with standardised scenarios according to the Guideline for Microscopic Evacuation Analysis (RiMEA)~\cite{kleinmeier-2019-cdyn}.

% Validation:
The validation of the transmission model, however, poses a challenge since empirical studies on local infection spread, in particular related to \sars{}, are scarce.
Fortunately, some data on superspreading events are available and are sufficiently detailed to be compared to simulations. Superspreading means that one or relatively few individuals infect numerous people \cite{lewis-2021-life}.
%entails an over-proportional transmission of COVID-19 \cite{lewis-2021-life}.
%that is,    
Since it also plays a significant role in the transmission dynamics of \sars{} \cite{althouse-2020-life}, it becomes the core of our validation.

%Many superspreading events have been recorded. 
Majra \ea\cite{majra-2021-life} presented an overview of recorded events. Unfortunately, only a few situations can be simulated with our model, and even fewer are suitable for validation. 
Firstly, the model is designed to capture the transmission via aerosol clouds mostly stagnant for at least several minutes. Thus, strong flows should not dominate the air circulation at the event.
Secondly, the event must not be too complex to be meaningfully represented in a simulation scenario. For example, the unknown routes of hundreds of guests at a carnival party would have to be guessed, making any comparison of data doubtful.

% Super spreading events
We find that the following events fit our purpose best:
\sars{} spread in a restaurant with ten infected people \cite{lu-2020-life} and
during a choir rehearsal, where $52$ of $61$ attendees became infected \cite{hamner-2020-life, miller-2020-life}.
%
% Limitation of reported cases
Both events occurred and were investigated in the early phase of the pandemic when science and society were still relatively unaware of \sars{}.
Consequently, other than the more recent events, 
measures such as physical distancing, air filtering or masks were absent.
These would introduce further complexity because they must be adequately modelled and validated. We avoid this for the validation reported here. Effects of these measures can be introduced into the model by adapting the values of parameters, \eg{} particles exhaled or half-life of the aerosol cloud.  
%
% Quantity of Interest
The reports  provide information about the number of secondary cases, \ie{} infected persons but not about the individual number of absorbed particles. 
We solve this by assuming that an agent is exposed if it inhales the same amount of pathogen as in the reference scenario, \ie{}, $3200$ virus particles. We will also compare to a dose of $1000$, as suggested in the literature.
The parameter settings for the locomotion model for both validation cases are listed in the appendix (\tref{tab:appxParams}). 

\begin{figure}[h]
	\centering
	\includegraphics[width=0.5\linewidth]{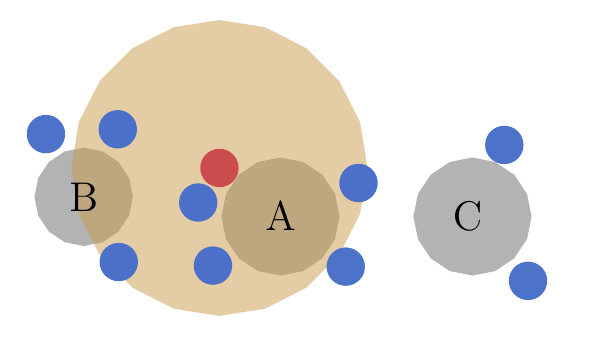}
	\caption{Model of restaurant topography, including tables and seats, according to the seating chart in \cite{lu-2020-life}. Susceptible (blue) agents sit in groups A, B and C around the tables (grey). The infectious (red) agent emits aerosol clouds (orange).
	}
	\label{fig:validationRestaurantScenario}
\end{figure}

% Restaurant scenario
We start with the spreading event in a restaurant in January 2020:  
Ten persons, divided into groups A, B and C, were sitting at adjacent tables. The infectious index patient belonged to group A. Group A shared the restaurant with group B for $\SI{53}{\minute}$ and with group C for $\SI{73}{\minute}$.
All ten individuals were tested positive after the restaurant visit. It was determined that the index case infected at least one member of groups B and C while in the restaurant. Further transmission among each group in the following days is considered possible. 
The topography is shown in \fref{fig:validationRestaurantScenario}.

%
%However, we neglect ventilation in the model which probably led to the infection among members of group C. Therefore, we expect less spatial spread in the model than in reality and, as a result, fewer than nine exposed agents in total and no exposed agent among group C.
%
We consider our simulation as qualitatively valid if we observe infection spread below the numbers in the study. We argue that the simulation ignores airflow between tables from the
restaurant's ventilation as well as any transmission after the event.
In the simulation, five agents become exposed, and among them, three from group A, two from group B and 
none from group C, \ie{}, half the number of the true cases. 

%We further elaborate on the simulation in \aref{appx:validationRestaurantScenario}.

% Choir practice
%Event / attendees
The choir rehearsal in March 2020 is significantly more complex: $1$ of $61$ attendees was symptomatic. After the practice, $33$ people, including the index patient, were tested positive. Twenty further attendees are considered probable cases because they became ill but were not tested. One person, initially classified as a probable case, tested negative after the onset of symptoms. Thus, we use a minimum of $32$ and a maximum of $52$ secondary cases as reference values to which we compare our simulation results.

% restriction / expectations
Again, we expect fewer exposed agents in our simulation than in reality. 
% - social interaction
In addition, we do not have sufficient information about close interactions between the attendees during arrival and departure to include them in the simulation. As a consequence, we ignore opportunities for droplet transmission.
% - droplet transmission 
Moreover, singing forcibly propels droplets, increasing pathogen spread.
% - possibly higher infectiousness of participants (older)
% Finally, our model assumes an averaged susceptibility of the population inspired by the close contact scenario even though the median age among both attendees and those who became ill was $69$ years.
% This may have an influence \cite{davies-2020-life}. 
% \srcomment{@Gesine: kennst du eine Quelle, mit der sich die Aussage (Ältere Menschen sind empfänglicher als jüngere) stützen lässt.}

\begin{figure}[h]
	\centering
	\includegraphics[width=\linewidth]{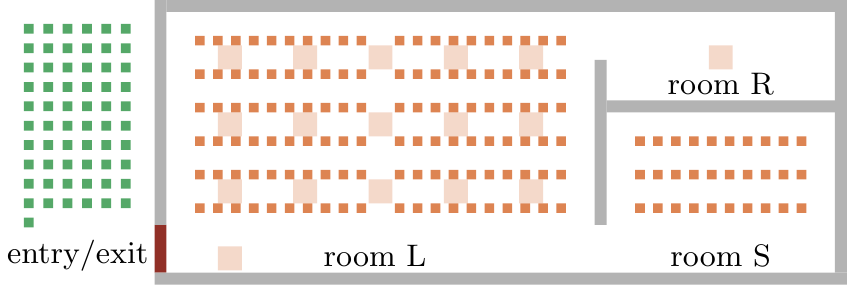}
	% Topographie Masse ohne Waende: room L ~18.4m x 11.9m, room S ~9.6m x 6.7m, gesamt ~28.5m x 11.9m
	\caption{Choir practice topography: a large (L) hall, a small (S) room and a restroom (R). Agents are spawned by sources (green), they approach an intermediate target, remain there for a given period and move to the next one (targets all orange). Small orange squares represent chairs. Large, light orange squares define positions where agents gather in small groups during the break. The agents exit the scenario through the red door.
	}
	\label{fig:validationChoirPracticeTopography}
\end{figure}

% Topography
The choir practice occurred in a large room and, partly, in a smaller room. The attendees changed their positions from time to time, which we model by allowing the agents to move from one intermediate target to another.
The study did not provide a seating chart citing privacy concerns.
%
% According to the authors, it is not essential because the attack rate is relatively high and the choir members changed seats during the practice.
%
Thus, we imitate the seating arrangement from \cite{miller-2020-life} qualitatively but must make assumptions about the floor plan: \fref{fig:validationChoirPracticeTopography}. The rooms cover an area of approximately $\SI{28}{\m} \times \SI{12}{\m}$. Smaller intermediate targets (orange) mark possible seating positions during the practice sessions.

% Schedule
Miller \ea\cite{miller-2020-life} report the following schedule: 
During the first $T_1 = \SI{45}{\minute}$, all choir members practised together in the large room L. Some seats between individuals were empty.
A $45$-min ($T_2$) sectional rehearsal followed during which the attendees were divided into two groups, one in room L, the other in room S. The index case remained in room L.
After that, there was a break of approximately $T_3 = \SI{10}{\minute}$. This allowed for mingling, and a few people, including the index case, used the restroom. The positions during the break are not reported. We assume that the choir members gathered in small groups distributed across room L.
For the last session of $T_4 = \SI{50}{\minute}$, everybody returned to their original positions in room L.

% Simplifications
We simplify the schedule. Firstly, we choose practice sessions of equal length ($T_1 = T_2 = T_4 = \SI{45}{\minute}$), which makes the definition of intermediate target positions in the simulation easier. Secondly, we ignore the fact that the attendees arranged chairs before and after the rehearsal, arguing that the time for this appears short compared to the entire practice.
While these two aspects may be negligible, we acknowledge that the attendees' exact positions in space and time would be essential. They would reveal where high pathogen concentrations can occur and, where exposure is likely. Unfortunately, this information was not recorded and must remain uncertain.

\begin{figure}[h]
	\centering
	\includegraphics[width=\linewidth]{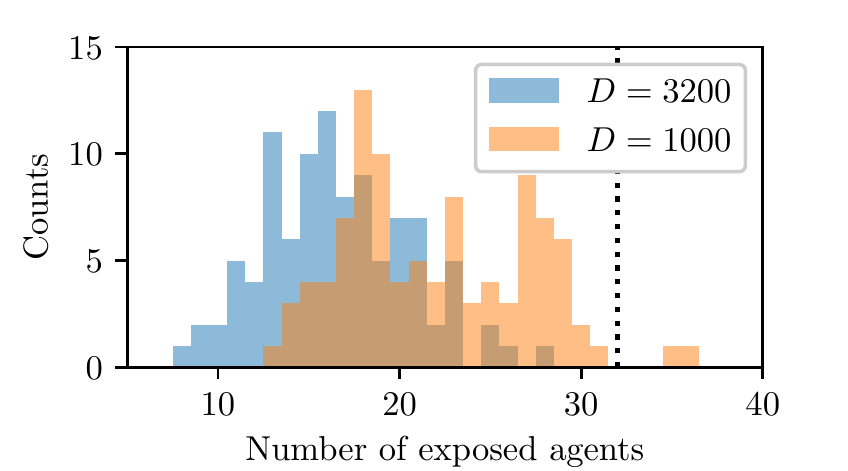}
	\caption{Number of exposed agents for $M=100$ simulations. Blue: infectious dose of $D=3200$ pathogen particles. Orange: a more vulnerable population with $D=1000$. The dotted line indicates the $32$ confirmed secondary cases
	of the true spreading event.}
	\label{fig:validationChoirPracticeHistogram}
\end{figure}

% Monte Carlo
We deal with this by applying Monte Carlo techniques: We evaluate the model $M=100$ times, collect the simulation output and summarise it statistically. The simulations differ only in the agents' paths. We achieve this by randomly mapping the agents to their intermediate targets for each simulation set-up.
\fref{fig:validationChoirPracticeHistogram} shows a histogram of the simulation output, \ie{}, the number of exposed agents.
As discussed in \sref{sec:Parameters}, the infectious dose is uncertain and may depend on individuals' immune systems. A dose of $D=3200$ pathogen particles corresponds to the close contact scenario, whereas $D=1000$ was estimated in \cite{popa-2020-life}.
  
We observe that the simulation results for both parameter choices are of the order of magnitude that was reported for the real event, but with fewer exposed agents. This is what we expected and regarded the simulation result as proof of our model's validity.

%
%
% *** alternative figure that shows histogram for two different cases (A: normal condition, B: fewer free chairs) ***
%\begin{figure}[h]
%	\centering
%	\includegraphics[width=\linewidth]{choirScenario/histInfectionStatus_AB.pdf}
%	\caption{We count the number of exposed agents for $N=100$ simulations. Case A represents the normal condition. Case B models the same scenario but the last two rows in room L are not occupied, \ie{} the agents sit closer together. In either case, the histogram shows that most simulations yield fewer exposed agents than $32$ confirmed secondary cases as reported in \cite{hamner-2020-life}.}
%	\label{fig:validationChoirPracticeHistogram}
%\end{figure}

% Implications

% sanity check in queue scenario
%In addition to the considered super spreading events, simulating the close contact scenario in \sref{sec:calibration} already serves as sanity check and thus contributes to the validation.
 
% (comparing InfectionModel (with short incubation period) with a "network SIR-model")
% (comparing model output with Wells-Riley output)

\section{Application to a queue scenario}  \label{sec:application}
\begin{figure}[ht]
\centering
\includegraphics[width=0.25\linewidth]{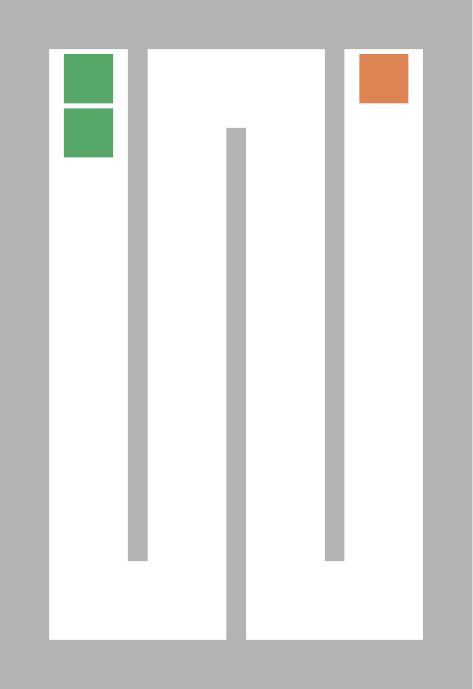}
\caption{Queue topography: agents (blue) start at the source (green) and move towards the service unit (orange). Belt barriers are shown as grey obstacles.}
\label{fig:CompModelExampleScenarioTopography}
\end{figure}

% Demonstration of the model based on the queueing scenario
% Intro, general description of the scenario
In this section, we use the transmission model to evaluate exposure in
a queue in front of a service unit, \eg{} a counter at a shopping centre, a cinema, an office or any other similar situation.
We simulate one highly infectious person among several susceptible pedestrians. 
In the green area, one source spawns nine susceptible agents, and a second source spawns a single infectious agent at the fifth position. Belt barriers guide the meandering queue to the counter, where each agent is served for a fixed time $T_s = \SI{120}{\s}$. The agents immediately leave the topography as soon as they pass the counter.
% Topography
The topography, as shown in \fref{fig:CompModelExampleScenarioTopography}, may represent a part of a building, covering an area of $\SI{5}{\m} \times \SI{7}{\m}$. 
%
% Parameter settings
In terms of validation scenarios, the parameter settings for the locomotion model are listed in \tref{tab:appxParams} of the appendix, allowing third parties to repeat and check our computer experiment.

% Queue
%The time spans between the agents' arrivals at the service unit follow approximately a uniform distribution.
%The inter-arrival times between agents and the service times are uniformly distributed. 
%The service time $T_s$ can be interpreted as the average time needed for the payment process at the counter.
%
%

\begin{figure}
\centering
\begin{subfigure}[b]{0.22\linewidth}
	\centering
	\includegraphics[width=\textwidth]{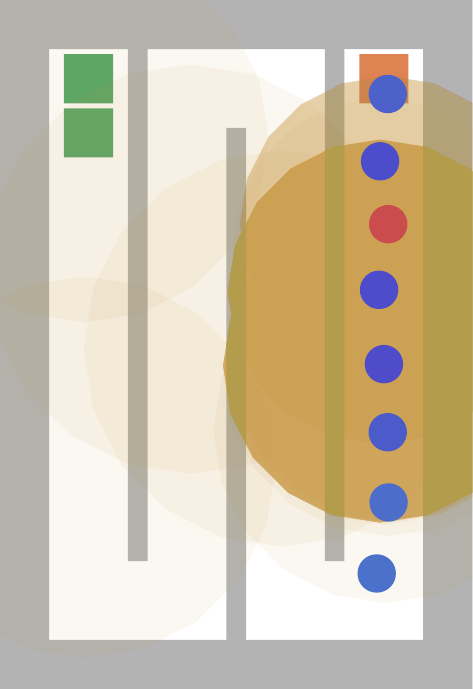}
    \caption{$t=\SI{300}{\s}$}
\end{subfigure}
\hfill
\begin{subfigure}[b]{0.22\linewidth}
	\centering
	\includegraphics[width=\textwidth]{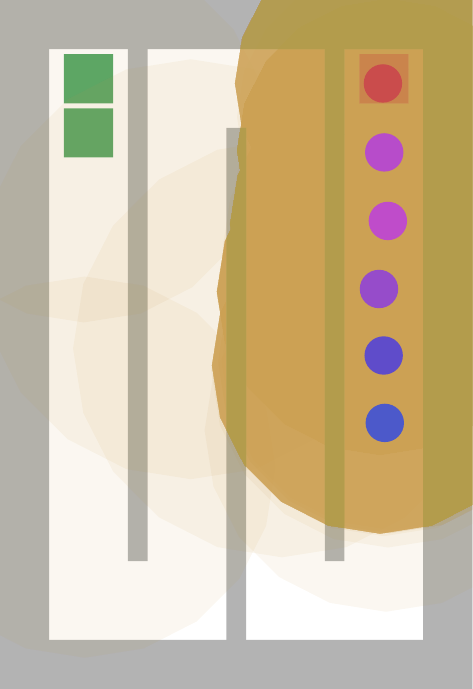}
    \caption{$t=\SI{600}{\s}$}
\end{subfigure}
\hfill
\begin{subfigure}[b]{0.22\linewidth}
	\centering
	\includegraphics[width=\textwidth]{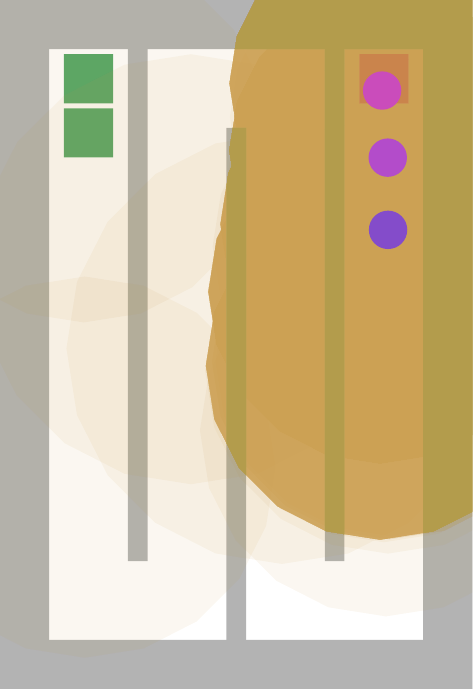}
    \caption{$t=\SI{900}{\s}$}
\end{subfigure}
\hfill
\begin{subfigure}[b]{0.25\linewidth}
	\centering
	\includegraphics[width=\textwidth]{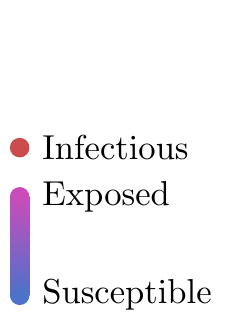} \vfill
    \caption*{}
\end{subfigure}
\caption{Susceptible agents (blue) are exposed to pathogens in aerosol clouds (orange circles) exhaled by an infectious agent (red). The opacity of aerosol clouds reflects their current pathogen concentrations. The agents' colour change, from blue to violet,
indicates their degree of exposure.}
\label{fig:CompModelExampleScenarioSimulation}
\end{figure}

% Simulation results
\fref{fig:CompModelExampleScenarioSimulation} shows the queue for several time steps. 
The colour, ranging from blue (susceptible) to violet (exposed), indicates the number of pathogens accumulated in an agent. 
%
% visualization of aerosol clouds
All aerosol clouds remain at their initial positions because we assume stable air layers. Hence, high concentrations occur where many clouds are superimposed. 

This pathogen concentration visualisation aids us in identifying potentially risky situations.
%, or more precisely, which positions in relation to the infectious agent's trajectory are problematic.
%
% qualitative comparison among agents within the scenario
The sixth, seventh and eighth positions in this queue, behind the infectious agent, are critical, as agents directly step into and remain in the recently contaminated area.  
%
% Evaluation of the queuing scenario relative to the reference scenario
The agents at these positions absorb more than $D = 3200$ pathogen particles, \ie{}, they carry at least the same infection risk as the close contact in the reference scenario.
However, the seventh and eighth positions are not a close contact position. The infectious agent queues for $\SI{10}{\minute}$. The seventh and eighth agent are within $\SI{1.5}{\m}$ of the infectious agent for less than half of this time and continue to inhale contaminated air after the infectious agent has left.

% implications / variations of the scenario
Our simulation supports the claim that queues, in stagnant air, pose a severe infection risk. 
However, how could the risk of exposure be reduced?
%
%We propose to play through different settings by varying the model parameters. For example, one can simulate typical measures taken during the COVID-19 pandemic such as physical distancing or wearing masks. We emphasize that varying in particular the medical parameters requires sound knowledge. Since the knowledge related to COVID-19 is yet limited, the simulation outcomes should predominantly be used to make comparisons to other simulations and to derive qualitative results.
We propose to evaluate measures by varying the model parameters that reflect these measures. For example, a mask worn by the infectious agent would reduce the number of pathogens. 
In addition, social distancing could be introduced. Organisers could strive to avoid this type of queue, \eg{} by handing out service numbers, or they could install overhead ventilation.

\section{Conclusion and outlook}  \label{sec:conclusion_outlook}
% Conclusion
% > recap introduction
%Modelling the transmission of \sars between individuals can help to better understand the risk of becoming infected in different everyday situations.
% > recap methods and materials
%
We complemented microscopic crowd simulation with a new model for the transmission of pathogens via small aerosol particles within Vadere, an open-source framework for simulating pedestrian dynamics.
% short story
Infectious agents exhale pathogens bound to aerosol clouds, whereas susceptible individuals absorb pathogens.% from the surrounding. 
We calibrated parameters to the transmission of \sars{} and re-enacted two superspreading events for which we obtained qualitatively plausible results. 

% > How to employ the model
As a result, we demonstrated how to evaluate the risk of exposure in everyday situations using our simulation model: we observed the number of pathogen particles absorbed by agents in a typical queue.  We compared the result to a reference value obtained from a benchmark scenario: a close contact situation acknowledged as high risk in the context of \sars{} by official health authorities.
As long as there is no consensus on the true infectious dose for \sars{}, we propose interpreting agents as exposed if they inhale as many viruses as in the reference scenario.
%
% comparison of agents from the same simultion
%Alternatively, we compared the degree of exposure between the agents within the same scenario. In this manner, one can visualise and identify critical or non-critical positions or areas.

% Outlook
% (scientific communication)
% improving the model
As a next step, we plan to refine the temporal and spatial spread of aerosols in the model.
%
% appyling the model to different scenarios
In addition, we propose analysing additional scenarios, including the variation of parameters, to evaluate the effectiveness of measures, to account for immunisation as well as an evolving virus.
%
% transfering the model
Beyond that, we hope that our model will be adapted, when the need arises, by other scientists to investigate future pandemics.

\section*{Acknowledgement}
We thank Dr Mareike M\"ahler, Dr Laura K\"unzer, Dr D\'{e}sir\'{e}e Dahmen, and Nele Clemen (Team HF) for support with literature research related to the COVID-19 pandemic.
We thank Dr Angelika Kneidl (accu:rate GmbH) and Prof.~Dr Christian Schwarzbauer (Munich University of Applied Sciences HM) for valuable discussions and Dr Andreas Wieser (Ludwig Maximilian University of Munich) for information about the infectiousness of COVID-19 patients.

\paragraph{Authors' contributions}
%\srcomment{Ggf. weitere Ergänzungen?}
Conceptualisation \SR{}, \MG{} and \GK{}; Data curation \SR{}; Formal analysis \SR{} and \MG{}; Funding acquisition \GK{}; Investigation \SR{} and \GH{}; Methodology \SR{} and \MG{}; Project administration \GK{}; Resources \GK{}; Software \SR{}; Supervision \GK{}; Validation \SR{}; Visualization \SR{}; Writing – original draft \SR{}; Writing – review \& editing \MG{}, \GK{} and \GH{}. All authors gave final approval for publication and agree to be held accountable for the work performed therein.

\paragraph{Funding}
%TODO maybe add financial support by HM?
\SR{}, \MG{}, \GK{}, and \GH{} are supported by the German Federal Ministry of Education and Research through the project CovidSim (grant no. 13N15662).
%This work was financially supported by the Munich University of Applied Sciences HM through the Open Access Publishing program.

\paragraph{Competing interest}
The authors declare that there is no conflict of interest.

% \paragraph{Ethics statement}

\paragraph{Data accessibility}
The datasets supporting this article have been uploaded 
%as part of the supplementary material
to \cite{vadere-2021-cdyn}.

\clearpage

% BibTeX users use
\bibliographystyle{unsrtnat-initials-truncatedauthorlist} % mathematics and physical science
%\bibliography{LifeSciences,CollectiveDynamics} % name your BibTeX data base
\bibliography{extracted} % file extracted with jabref from LifeSciences.bib, CollectiveDynamics.bib

\appendix
\section{Detailed parameter set-up}
To allow third parties to replicate our computer experiment, we provide more detailed information about the model set-up.
In principle, all scenario files can be accessed on Gitlab \cite{vadere-2021-cdyn}.
However, the following parameter sets should be used to reproduce the simulations independently:
All (pseudo-)random numbers used in the simulations with Vadere can be generated with the seeds listed in \tref{tab:appxSeeds}.
\begin{table}[h!]
\caption{Simulation parameters.}
\label{tab:appxSeeds}
\begin{tabularx}{\linewidth}{X r}
\hline
\textbf{Scenario} & \code{fixedSeed} \\ \hline \hline
Close contact & $8838372581797678424$ \\ \hline
Restaurant & $4889043484410943750$ \\ \hline
Choir rehearsal & $-2054058476485033808$ \\ \hline
Queue & $1436250873317888407$ \\ \hline
\end{tabularx}
\end{table}

We also adapted the parameters of Vadere for the OSM to fit the agents' locomotion behaviour to the simulated situation. The modifications to the default settings are listed in \tref{tab:appxParams}.
The scenarios used for the validation require minor adaptations to the pedestrians' potential. These affect the distance between an agent and other agents or obstacles. 
For the queue scenario, we enabled a greater sensitivity regarding thin obstacles (belt barriers). We also reduced the agents' attraction towards their target (counter) to reach more realistic distances between the agents in the queue.
\begin{table*}[t]
\caption{Parameters of the Optimal Steps Model defining the agents' locomotion behaviour.}
\label{tab:appxParams}
\begin{tabularx}{\linewidth}{X l l l l}
\hline
\textbf{Scenario} & \code{seeSmallWalls} & \code{pedPotentialHeight} & \code{obstPotentialWidth} & \code{targetAttractionStrength} \\ \hline \hline
Default/close contact & \code{false} & $50$ & $0.8$ & $1$ \\ \hline
%Close contact & & & & \\ \hline
Restaurant & & $5$ & $0$ &  \\ \hline
Choir rehearsal & & $5$ & &  \\ \hline
Queue & \code{true} & & & $0.4$ \\ \hline
\end{tabularx}
\end{table*}

\end{document}